\documentclass[10pt,conference]{IEEEtran}
\IEEEoverridecommandlockouts

\usepackage{cite}
\ifCLASSINFOpdf
  \usepackage[pdftex]{graphicx,color}
  \graphicspath{{../pdf/}{../jpeg/}}
  \DeclareGraphicsExtensions{.pdf,.jpeg,.png}
\else
  \usepackage[dvips]{graphicx}
      \usepackage{color}

  \graphicspath{{../eps/}}
   \DeclareGraphicsExtensions{.eps}
\fi
\usepackage[cmex10]{amsmath}
\usepackage{amssymb}

\begin{document}
\title{On the Stability Region of Amplify-and-Forward Cooperative Relay Networks}

\author{\IEEEauthorblockN{Jubin Jose\IEEEauthorrefmark{1},
Lei Ying\IEEEauthorrefmark{2},
Sriram Vishwanath\IEEEauthorrefmark{1}}
\IEEEauthorblockA{\IEEEauthorrefmark{1}
Dept. of Electrical and Computer Engineering, The University of Texas at Austin\\
Email: \{jjose,sriram\}@ece.utexas.edu}
\IEEEauthorblockA{\IEEEauthorrefmark{2}Dept. of Electrical and Computer Engineering, Iowa State University\\
Email: leiying@iastate.edu
}
\thanks{\IEEEauthorrefmark{1} The authors were supported in part by the DoD, the ARO Young Investigator Program (YIP) and the NSF grant CNS-0831756.}}
\maketitle



\begin{abstract}
This paper considers an amplify-and-forward relay network with fading states. Amplify-and-forward scheme (along with its variations) is the core mechanism for enabling cooperative communication in wireless networks, and hence understanding the network stability region under amplify-and-forward scheme is very important. However, in a relay network employing amplify-and-forward, the interaction between nodes is described in terms of real-valued ``packets'' (signals) instead of discrete packets (bits). This restrains the relay nodes from re-encoding the packets at desired rates. Hence, the stability analysis for relay networks employing amplify-and-forward scheme is by no means a straightforward extension of that in packet-based networks. In this paper, the stability region of a four-node relay network is characterized, and a simple throughput optimal algorithm with joint scheduling and rate allocation is proposed.

\end{abstract}


\section{Introduction}

Relaying is central to wireless mesh and ad hoc networks, and is a potential enhancement for existing cellular networks. There is no ``wireless network'' without multi-hop communication, and thus, understanding the role and impact of different relaying techniques on wireless networks is of critical importance. There are multiple possible forwarding strategies that could be used by each relay node in the network, with the most popular ones being decode-and-forward, amplify-and-forward and quantize-and-forward. The forwarding strategy used at the relays and the resulting network capacity are inherently coupled, and it is not obvious which of the above forwarding strategies, if any, is information theoretically optimal for a particular network topology. For a few settings, amplify-and-forward has been shown to be better than decode-and-forward \cite{J:ChenAL08}, and for a few others, partial decode-and-forward has been shown to be optimal \cite{J:GA82}.

Other than purely information-theoretic rate calculations, there are practical reasons for the use of amplify/quantize-and-forward strategy. Decoding a packet at the relay requires an entire receive chain (demodulation and decoding) to be implemented at the relay, along with an entire transmit chain (re-encode, remodulate and retransmit). An amplify/quantize-and-forward relay can bypass this chain, processing the received signal directly to obtain the relay output. This simplification greatly impacts the cost, energy usage and size of the relays, and therefore, it is not surprising that the majority of the relays used in practice today are based on amplify-and-forward strategy. Also, in practice, amplify-and-forward is at par with decode-and-forward strategy with reduced complexity in processing \cite{C:YL05}. Therefore, it is imperative to understand the stability region of relay networks employing amplify-and-forward scheme.

For non-cooperative relaying, there is a vast body of literature on optimal resource allocation in wireless networks \cite{J:TE92,LinShrSri_06,B:GNT06} (and references therein). For cooperative relaying \cite{SE04,Thesis:S01}, the stability region of relay networks using decode-and-forward transmission has been studied in \cite{C:YB05,C:YB07}. In the decode-and-forward paradigm, the queues still consist of conventional \emph{decoded data} packets. In an amplify-and-forward relay network, each relay, however, observes an input sequence, whose ``rate'' is dependent on the state it was encoded for, which is not a conventional packet. There are three major differences between the amplify-and-forward relaying and conventional node-forwarding:
\begin{enumerate}
\item Each node in the network stores and ultimately forwards a real valued ``packet''. This is accomplished by quantizing the received analog signal to some finite precision and storing it in a buffer. This stored signal is later retransmitted without further processing.
\item The ``rate'' or effective bits per symbol of each real-valued ``packet'' is not the same, but variable depending on the encoding. 
\item A high-rate real-valued ``packet'' can only be forwarded when the channel can support the corresponding rate, i.e., the channel state should \emph{match} the ``packet'' rate, which is not required in the decode-and-forward scheme.
\end{enumerate}


We consider a four node relay network in this paper, and our main contributions include:
\begin{enumerate}
\item We introduce a new queue-architecture to store the input sequences, which allows us to optimally exploit different states to do amplify-and-forward. We characterize the stability region of the four-node relay network under the amplify-and-forward, which in general is larger than that under the conventional amplify-and-forward scheme which does not allow signal buffering.
\item We also propose a throughput-optimal algorithm, which achieves the stability region. An interesting and important property of our stabilizing algorithm is that it does not require the knowledge of the underlying distribution of the fading states.
\end{enumerate}

The rest of this paper is organized as follows. The next section introduces the system model. Section \ref{sec:rate} provides an information-theoretic achievable rate for amplify-and-forward relaying. Section \ref{sec:main} presents the algorithm for throughput-optimal stable cooperative relaying over wireless networks for the four node example. The paper concludes with Section \ref{sec:conclude}. The proofs are stated in the Appendix.

\section{System Model}

\begin{figure}[!t]
\centering
\scalebox{0.5}{\input{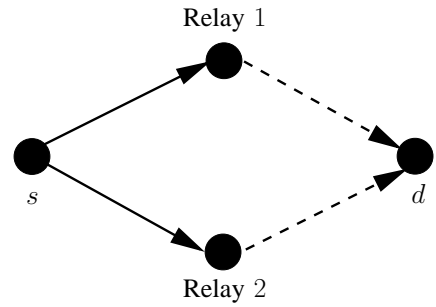}}
\caption{A Simple Cooperative Relay Network}
\label{fig:model}
\end{figure}

Our system model consists of a source, a destination and two relays as shown in Figure \ref{fig:model}. The relays $1$ and $2$ assist the source $s$ in transmitting to the destination $d$ through amplify-and-forward relaying. There is no direct link between the source and the destination. In addition, there is link activation constraint such that the links from the source to the relays and the links from the relays to the destination cannot be active at the same time. This constraint arises from the following two system limitations: \emph{(i)} a relay cannot receive and transmit simultaneously due to hardware limitations, and \emph{(ii)} one of the relays (say relay $1$) cannot receive from the source at the same time as the other relay (relay $2$) is transmitting to the destination. In essence, we assume that the transmissions of the two relays heavily interfere with each other. We consider a discrete time model for data transmission over the links. All the links undergo slow fading.  Further, we make the following assumptions:
\begin{enumerate}
\item[(1)] Assumption 1: There is coordination between the relays, i.e., they transmit signals synchronously.
\item[(2)] Assumption 2: The fading process is independent and identically distributed (i.i.d.) block fading with block length of $T$ symbols. Without loss of generality, we assume $T=1.$
\item[(3)] Assumption 3:  There is no power control over states. We consider an average power constraint of $P$ per block per node in the network, and additive Gaussian noise of unit variance at each receiver in the network.
\end{enumerate}


To simplify notation, we denote the relays by $n, n\in\{1,2\}$, the wireless link between the source and the relay $n$ by $l_{sn}$, and the wireless link between the relay $n$ and the destination by $l_{nd}$. Let the fading state of link $l_{sn}$ be $F_{sn}$, and the fading state of link $l_{nd}$ be $F_{nd}$. We assume that $F_{sn}$  and $F_{nd}$ are discrete non-negative random variables which take values from ${\cal{F}}$. In every block, let the probability of the fading state ${\mathbf{f}}=[f_{s1}, f_{s2}, f_{1d}, f_{2d}] \in {\cal{F}}^4$ be $\pi_{\mathbf{f}}$.


Consider a given fading state $\mathbf{f}$. If the source transmits, then the received signals at the relays are
\begin{equation}
\label{y:relay}
y_n =  \sqrt{f_{sn}} x_s + w_n,  \hbox{ for } n=1, 2,
\end{equation}
where $x_s$ denotes the symbol transmitted from the source, and $y_n$ denotes the symbol received at the relay $n$. If the relays transmit, then the received signal at the destination is
\begin{equation}
\label{y:dest}
y_d =  \sum_{n=1}^2 \sqrt{f_{nd}} x_n + w_d, 
\end{equation}
where $x_n$ denotes the symbol transmitted from the relay $n$, and $y_d$ denotes the symbol received at the destination. Here, $w_n$ and $w_d$ are i.i.d. zero-mean additive Gaussian noise of unit variance at the relay $n$ and the destination.




\section{Amplify-and-Forward Achievable Rate}
\label{sec:rate}

In the static case without any link activation constraint, amplify-and-forward commonly refers to the relaying scheme at the relays that transmit (in every time slot) scaled versions of the received signals in the previous time slot. The scaling parameters at the relays are determined as a function of the signal-to-noise ratios (SNRs). These scaling parameters should also ensure that the average power constraints at the relays are satisfied. For the setting explained in the previous section, we look at an amplify-and-forward scheme (denoted by AF scheme) in which the relays can transmit any of the previously received signal vectors or choose not to transmit. We assume that any received signal vector at the relays can be transmitted to the destination only once.

We say that a symbol $x_s$ is transmitted by the source to the destination over state $\mathbf{g}=[g_{s1}, g_{s2}, g_{1d}, g_{2d}]\in {\cal F}^4$, if the source transmits during a fading state of the form $[g_{s1}, g_{s2}, *, *] $ and the relays transmit during a fading state of the form $[*, *, g_{1d}, g_{2d}]$. Consider a symbol $x_s$ transmitted by the source to the destination over some state $\mathbf{g}$. Let the average power used at the source be $P_s^{\mathbf{g}}$ and at the relay $n$ be $P_n^{\mathbf{g}}$. These parameters are later optimized for the state $\mathbf{g}$. From (\ref{y:relay}), (\ref{y:dest}), the received symbol at the destination is
\begin{equation}
\label{net_ch}
y_d = \sum_{n=1}^2 \left[ \sqrt{\frac{g_{nd}P_n^{\mathbf{g}}}{g_{sn}P_s^{\mathbf{g}}+1} } \left(\sqrt{g_{sn}} x_s + w_n \right) \right]+ w_d,
\end{equation}
where $x_s$ has zero mean and variance $P_s^{\mathbf{g}}$. From (\ref{net_ch}), it is straightforward to see that the maximum rate we can obtain is
\begin{equation}
\label{rkl}
r_{\mathbf{g}}= \max_{P_s^{\mathbf{g}}, P_1^{\mathbf{g}}, P_2^{\mathbf{g}} \le P} C\left( \frac{P_s^{\mathbf{g}}\left(\sqrt{g_{s1}c_1}+\sqrt{g_{s2}c_2}\right)^2}{c_1+c_2+1}\right),
\end{equation}
where $c_n={(g_{nd}P_n^{\mathbf{g}})}/{(g_{sn}P_s^{\mathbf{g}}+1)}$ and $C(x)=\frac{1}{2}\log_2(1+x).$

\newtheorem{remark}{\hspace{-12pt} Remark}
\newtheorem{lemma}{\hspace{-12pt} Lemma}
\newtheorem{theorem}{\hspace{-12pt} Theorem}

\begin{remark}
The rate in (\ref{rkl}) is equal to the maximum achievable rate using amplify-and-forward scheme in the static case with fixed channel state $\mathbf{g}$ and full-duplex constraint \cite{Thesis:S01}.
\end{remark}

\begin{remark}
The power optimization in (\ref{rkl}) will result in utilizing maximum power at the source and one of the relays. In general, the other relay will result in using
lower power than the maximum available power\cite{Thesis:S01}.
\end{remark}

For handling variable rate allocation, we maintain separate virtual queues at each relay based on the possible rates of the real-valued ``packet''. This is necessitated by the fact that encoding and decoding in amplify-and-forward relaying is an end-to-end process. Thus, variable rate coding at the source depending on the state must be done to ensure that: \emph{(i)} the virtual queues at the relays are stabilized, and \emph{(ii)} the relays possess sufficient ``packets'' at variable rates for throughput optimal operation. We denote by ${Q}_{n}^{\mathbf{g}}[t]$ the length of the virtual queue maintained for state $\mathbf{g}$, where $\mathbf{g}=[g_{s1}, g_{s2}, g_{1d}, g_{2d}]\in {\cal F}^4.$ Note that when the system is in fading state $\mathbf{f},$ for any state $\mathbf{g}$ such that $g_{s1}=f_{s1}$ and $g_{s2}=f_{s2},$ a ``packet'' generated with rate $r_{\mathbf{g}}$ can be successfully transmitted from the source to the relays; and for any state $\mathbf{g}$ such that $g_{1d}=f_{1d}$ and $g_{2d}=f_{2d},$ a ``packet'' generated with rate $r_{\mathbf{g}}$ can be successfully transmitted from the relays to the destination. We define $I_s=\{(\mathbf{f},\mathbf{g})|g_{s1}=f_{s1}, g_{s2}=f_{s2}\}$, and $I_d=\{(\mathbf{f},\mathbf{g})|g_{1d}=f_{1d}, g_{2d}=f_{2d}\}$.

The throughput region of the network is defined in the following lemma.

\begin{lemma}
\label{lem:r}
A rate $r$ is supportable in the four-node relay network using an AF scheme only if there exist $a_{\mathbf{f}}^{\mathbf{g}}\geq 0$ and $b_{\mathbf{f}}^{\mathbf{g}}\geq 0$ such that
\begin{eqnarray}
&\displaystyle r=\sum_{\mathbf{g}\in{\cal F}^4}\sum_{\mathbf{f}\in{\cal F}^4} \left( \pi_{\mathbf{f}} a_{\mathbf{f}}^{\mathbf{g}} r_{\mathbf{g}} \mathbf{1}_{\{(\mathbf{f},\mathbf{g})\in I_s\}} \right),\label{eq: fc1}\\
&\displaystyle \sum_{\mathbf{f}\in{\cal F}^4} \pi_{\mathbf{f}} a_{\mathbf{f}}^{\mathbf{g}} \mathbf{1}_{\{(\mathbf{f},\mathbf{g})\in I_s\}}=\sum_{\mathbf{f}\in{\cal F}^4} \pi_{\mathbf{f}} b_{\mathbf{f}}^{\mathbf{g}}\mathbf{1}_{\{(\mathbf{f},\mathbf{g})\in I_d\}},   &\forall \mathbf{g},\label{eq: fc2}\\
&\displaystyle \sum_{\mathbf{g}\in {\cal F}^4}a_{\mathbf{f}}^{\mathbf{g}}+b_{\mathbf{f}}^{\mathbf{g}} \leq 1, &\forall \mathbf{f},
\label{eq: overall}
\end{eqnarray} where $\pi_{\mathbf{f}}$ is the probability that the system is in fading state $\mathbf{f}$ and $\mathbf{1}_{\{E\}}$ denotes the indicator function of event $E$. When the system is in fading state $\mathbf{f},$ the source can transmit one ``packet'' to any one virtual queue $\mathbf{g}$ for which $\mathbf{1}_{\{(\mathbf{f},\mathbf{g})\in I_s\}}=1;$ or the relays can transmit one ``packet'' in any one virtual queue $\tilde{\mathbf{g}}$ to the destination for which $\mathbf{1}_{\{(\mathbf{f},\mathbf{\tilde{g}})\in I_d\}}=1.$
\end{lemma}

\begin{IEEEproof}[\hspace{-21pt} Proof]
Let $a_{\mathbf{f}}^{\mathbf{g}}$ be the fraction of time the source transmits to virtual queue $\mathbf{g}$ when the system is in fading state $\mathbf{f},$ and  $b_{\mathbf{f}}^{\mathbf{g}}$ be the fraction of time the relays transmit the ``packets'' in virtual queue $\mathbf{g}$ to the destination when the system is in fading state $\mathbf{f}$. Hence, Inequality (\ref{eq: overall}) says that the overall fraction should be no more than one. Equality (\ref{eq: fc1}) is the flow conservation constrain for the source to all the virtual queues, and Equality (\ref{eq: fc2}) is the flow conservation constrain for each individual virtual queue. The detailed proof is given in the Appendix.
\end{IEEEproof}

\begin{figure}[!t]
\centering
\includegraphics[width=70mm]{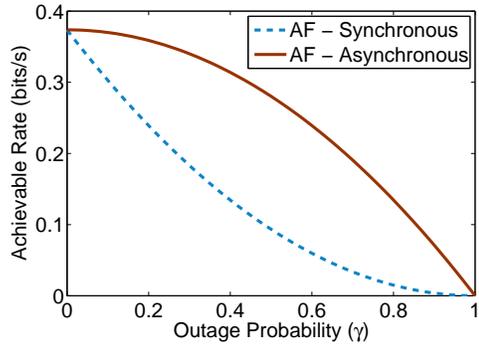}
\caption{Amplify-and-Forward: Asynchronous vs Synchronous}
\label{fig:comp}
\end{figure}

We can obtain rates \emph{strictly} greater than the average of rates over all fading states by asynchronously combining states between the source and the relays and the relays and the destination. We will demonstrate this using a simple example. Let $P = 1$, ${\cal{F}}=\{0,1,10\}$. Consider the fading distribution such that fading states $[0 , 0 , 0 , 0]$, $[0 , 0 , 10 , 10]$, $[1 , 1 , 0 ,0]$ and $[1 , 1 , 10 , 10]$ occur with probabilities $\gamma^2$, $\gamma \bar{\gamma}$, $\bar{\gamma}\gamma$, and $\bar{\gamma}^2$, respectively. Here, $\bar{\gamma}=1-\gamma$ and $\gamma$ can be considered as the probability of outage in this example. The rate corresponding to state $[1 , 1 , 10 , 10]$ alone is non-zero, which is $C(20/11)$. In this example, it is easy to observe that we can achieve $(0.5\bar{\gamma}^2+\gamma\bar{\gamma})C(20/11)$ bits/transmission with asynchronous combining of states whereas $0.5\bar{\gamma}^2 C(20/11)$ bits/transmission is the average of rates over different fading states. These achievable rates are plotted in Figure \ref{fig:comp}.

\section{Throughput Optimal Algorithm and Stability Analysis}
\label{sec:main}

We consider an i.i.d. arrival process $A[t]$ for the data bits at the source $s$ with mean rate $\lambda$ and bounded variance. Let the queue at the source be $Q_s$ with queue length $Q_s[t]$ at time $t$. At the relay $n$, we assume $|{\cal F}^4|$ different virtual queues with queue length $Q_{n}^{\mathbf{g}}[t],$ which is the virtual queue length corresponding to state $\mathbf{g}$ at time $t.$ The queue $Q_s$ consists of bits whereas queue $Q_{n}^{\mathbf{g}}$ consists of real-valued ``packets'' (quantized to required precision) encoded for state $\mathbf{g}$ at rate $r_{\mathbf{g}}$. We will show that this queue-architecture at the relays can be utilized to obtain a throughput-optimal algorithm.


In this section, we provide an algorithm, based on back-pressure \cite{J:TE92} which is throughput-optimal, and that does not require the knowledge of the fading distribution. The algorithm has similarities with the maximum differential backlog algorithm for conventional networks and its generalization to cooperative relaying with decode-and-forward in \cite{C:YB07}. However, the fact that ``packets'' at the relays have variable rate introduces different weighting factors for different ``packets''.

\noindent {\bf Back-pressure-based amplify-and-forward algorithm:} Suppose the system is in fading state $\mathbf{f}[t]$ at time slot $t,$ then the central coordinator computes $$\mathbf{g}^*_1\in\arg\max_{\mathbf{g}} \left(Q_s[t] - r_{\mathbf{g}} \sum_{n=1}^2Q_{n}^{\mathbf{g}}[t]\right)r_{\mathbf{g}}\mathbf{1}_{\{(\mathbf{f},\mathbf{g})\in I_s\}}$$ and $$\mathbf{g}^*_2\in\arg\max_{\mathbf{g}} \left(r_{\mathbf{g}} \sum_{n=1}^2Q_{n}^{\mathbf{g}}[t]\right) r_{\mathbf{g}}\mathbf{1}_{\{(\mathbf{f},\mathbf{g})\in I_d\}}.$$ If $$\left(Q_s[t] - r_{\mathbf{g}^*_1} \sum_{n=1}^2Q_{n}^{\mathbf{g}^*_1}[t]\right)r_{\mathbf{g}^*_1}\geq \left(r_{\mathbf{g}^*_2} \sum_{n=1}^2Q_{n}^{\mathbf{g}^*_2}[t]\right) r_{\mathbf{g}^*_2},$$ then the system operates the links from the source to the relays and transmits $r_{\mathbf{g}^*_1}$ bits, encoded into one real-valued ``packet'', from queue $Q_s[t]$ to queues $Q_{1}^{\mathbf{g}^*_1}[t]$ and $Q_{2}^{\mathbf{g}^*_1}[t].$ Otherwise, the system operates the links from the relays to the destination and transmits one real-valued ``packet'' from queues $Q_{1}^{\mathbf{g}^*_2}[t]$ and $Q_{2}^{\mathbf{g}^*_2}[t]$ to the destination.
\rightline{$\IEEEQED$}

\begin{theorem}
\label{thm}
The back-pressure-based amplify-and-forward algorithm stochastically stabilizes the queues for any $\lambda$ if there exists $\epsilon>0$ such that $\lambda+\epsilon$ is within the throughput region given in Lemma \ref{lem:r}. Hence, the throughput region in Lemma \ref{lem:r} is also the stability region of the four-node relay network.
\end{theorem}
\begin{IEEEproof}[\hspace{-21pt} Proof] The proof is provided in the Appendix. 
\end{IEEEproof}

In the above algorithm, the weight associated with the virtual queue for the ``packets'' transmitted from the source to the relays in state $\mathbf{g}$ is $\left(Q_s[t] - r_{\mathbf{g}} \sum_{n=1}^2Q_{n}^{\mathbf{g}}[t]\right)\mathbf{1}_{\{(\mathbf{f},\mathbf{g})\in I_s\}}$. The sum $\sum_{n=1}^2Q_{n}^{\mathbf{g}}[t]$ appears as the two relays cooperate to transmit the ``packets''. This sum has a normalization factor of $r_{\mathbf{g}}$ as each ``packet'' corresponds to $r_{\mathbf{g}}$ effective bits. The term $\mathbf{1}_{\{(\mathbf{f},\mathbf{g})\in I_s\}}$ arises as, given a fading state, only ``packets'' corresponding to certain states can be transmitted from the source to the relays. Similar comments apply to the weight associated with the virtual queue for the ``packets'' transmitted from the relays to the destination in state $\mathbf{g}$.

\section{Conclusion}
\label{sec:conclude}
In this paper, we characterize the maximum stable throughput for a two-hop cooperative relay network. A key feature of this paper is that we bring together  physical and network layer constraints in characterizing this throughput. We believe that the analysis conducted in this work and the algorithm can be generalized to other cooperative relay networks employing an amplify-and-forward strategy. We also believe that this framework can be used to analyze other forwarding strategies including partial-decode-and-forward and quantize-and-forward.



\appendix


\subsection{Proof of Lemma \ref{lem:r}}
Let $a_{\mathbf{f}}^{\mathbf{g}}$ be the fraction of time the source transmits ``packets'' corresponding to state $\mathbf{g}$ when the system is in fading state $\mathbf{f},$ and  $b_{\mathbf{f}}^{\mathbf{g}}$ be the fraction of time the relays transmit the ``packets'' corresponding to state $\mathbf{g}$ when the system is in fading state $\mathbf{f}$. It is clear that the maximum achievable rate $r_{max}$ using the AF scheme is
\begin{eqnarray}
\label{max_rate}
& \max \limits_{a_{\mathbf{f}}^{\mathbf{g}}, b_{\mathbf{f}}^{\mathbf{g}}} & \sum_{\mathbf{g}\in{\cal F}^4}\min \left\{\sum_{\mathbf{f}\in{\cal F}^4} \pi_{\mathbf{f}} a_{\mathbf{f}}^{\mathbf{g}}r_{\mathbf{g}},\sum_{\mathbf{f}\in{\cal F}^4} \pi_{\mathbf{f}} b_{\mathbf{f}}^{\mathbf{g}}r_{\mathbf{g}} \right\},   \\
& \text{s.t.} & \sum_{\mathbf{g}\in{\cal F}^4} \left(a_{\mathbf{f}}^{\mathbf{g}} + b_{\mathbf{f}}^{\mathbf{g}}\right) \le 1, \quad \forall \mathbf{f}, \nonumber \\
& & a_{\mathbf{f}}^{\mathbf{g}} = 0, \quad \forall (\mathbf{f},\mathbf{g}) \notin I_s, \nonumber \\
& & b_{\mathbf{f}}^{\mathbf{g}} = 0, \quad \forall (\mathbf{f},\mathbf{g}) \notin I_d, \nonumber \\
& & a_{\mathbf{f}}^{\mathbf{g}}, b_{\mathbf{f}}^{\mathbf{g}} \geq 0, \quad \forall \mathbf{f,g}. \nonumber
\end{eqnarray}

Now, we will prove that for any $r \le r_{max}$ there exists $a_{\mathbf{f}}^{\mathbf{g}}\geq 0$ and $b_{\mathbf{f}}^{\mathbf{g}}\geq 0$ such that (\ref{eq: fc1}), (\ref{eq: fc2}), and (\ref{eq: overall}) are satisfied. Let $\hat{a}_{\mathbf{f}}^{\mathbf{g}}, \hat{b}_{\mathbf{f}}^{\mathbf{g}}$ be an optimal solution to the problem (\ref{max_rate}). Consider the following assignment. For each state $\mathbf{g}$, let $a_{\mathbf{f}}^{\mathbf{g}} = ({r}/r_{max})\theta_{\mathbf{g}}\hat{a}_{\mathbf{f}}^{\mathbf{g}}$, and $b_{\mathbf{f}}^{\mathbf{g}} = ({r}/r_{max})\eta_{\mathbf{g}}\hat{b}_{\mathbf{f}}^{\mathbf{g}}$. Choose $\theta_{\mathbf{g}}$ and $\eta_{\mathbf{g}}$ such that $\sum_{\mathbf{f}\in{\cal F}^4} \pi_{\mathbf{f}} a_{\mathbf{f}}^{\mathbf{g}}r_{\mathbf{g}} = \sum_{\mathbf{f}\in{\cal F}^4} \pi_{\mathbf{f}} b_{\mathbf{f}}^{\mathbf{g}}r_{\mathbf{g}}=({r}/r_{max})\min \left\{\sum_{\mathbf{f}\in{\cal F}^4} \pi_{\mathbf{f}} \hat{a}_{\mathbf{f}}^{\mathbf{g}}r_{\mathbf{g}},\sum_{\mathbf{f}\in{\cal F}^4} \pi_{\mathbf{f}} \hat{b}_{\mathbf{f}}^{\mathbf{g}}r_{\mathbf{g}} \right\}$. Note that $0 \le \theta_{\mathbf{g}} \le 1$ and $0\le \eta_{\mathbf{g}} \le 1$. Therefore, this assignment satisfies (\ref{eq: fc1}), (\ref{eq: fc2}), and (\ref{eq: overall}). This completes the proof.

\subsection{Proof of Theorem \ref{thm}}

Consider the following optimization problem:
\begin{eqnarray}
\label{max_rate_eq}
& \max \limits_{a_{\mathbf{f}}^{\mathbf{g}}, b_{\mathbf{f}}^{\mathbf{g}}} & \sum_{\mathbf{g}\in{\cal F}^4} \sum_{\mathbf{f}\in{\cal F}^4} \pi_{\mathbf{f}} a_{\mathbf{f}}^{\mathbf{g}}r_{\mathbf{g}},  \\
& \text{s.t.} &\sum_{\mathbf{f}\in{\cal F}^4} \pi_{\mathbf{f}} a_{\mathbf{f}}^{\mathbf{g}}r_{\mathbf{g}} \le \sum_{\mathbf{f}\in{\cal F}^4} \pi_{\mathbf{f}} b_{\mathbf{f}}^{\mathbf{g}}r_{\mathbf{g}}, \quad \forall \mathbf{g}, \nonumber \\
&& \sum_{\mathbf{g}\in{\cal F}^4} \left(a_{\mathbf{f}}^{\mathbf{g}} + b_{\mathbf{f}}^{\mathbf{g}}\right) \le 1, \quad \forall \mathbf{f}, \nonumber \\
& & a_{\mathbf{f}}^{\mathbf{g}} = 0, \quad \forall (\mathbf{f},\mathbf{g}) \notin I_s, \nonumber \\
& & b_{\mathbf{f}}^{\mathbf{g}} = 0, \quad \forall (\mathbf{f},\mathbf{g}) \notin I_d, \nonumber \\
& & a_{\mathbf{f}}^{\mathbf{g}}, b_{\mathbf{f}}^{\mathbf{g}} \geq 0, \quad \forall \mathbf{f,g}. \nonumber
\end{eqnarray}
Let the optimal objective value of problem (\ref{max_rate_eq}) be $\bar{r}_{max}$. We will show that $\bar{r}_{max}=r_{max}$. Since any feasible assignment for (\ref{max_rate_eq}) is a feasible assignment for (\ref{max_rate}) and has the same objective value, $\bar{r}_{max} \le r_{max}$. Since any feasible assignment for (\ref{max_rate}) lead to another feasible assignment for (\ref{max_rate_eq}) with same objective value (as shown in the the proof of Lemma \ref{lem:r}), $r_{max} \le \bar{r}_{max}$. Hence,  $\bar{r}_{max}=r_{max}$. The structure of this alternate characterization is used later in the proof.

When the fading state at time $t$ is $\mathbf{f}$, the algorithm in Section \ref{sec:main} is based on the optimization problem:
\begin{eqnarray}
\label{back_pressure}
& \max \limits_{\alpha_{\mathbf{f}}^{\mathbf{g}} , \beta_{\mathbf{f}}^{\mathbf{g}}} & \sum_{\mathbf{g}\in{\cal F}^4} \left(Q_s[t] - r_{\mathbf{g}} \sum_{n=1}^{2}Q_n^{\mathbf{g}}[t]\right) r_{\mathbf{g}} \alpha_{\mathbf{f}}^{\mathbf{g}} + \\
& & \sum_{\mathbf{g}\in{\cal F}^4} \left(r_{\mathbf{g}} \sum_{n=1}^{2}Q_n^{\mathbf{g}}[t]\right) r_{\mathbf{g}} \beta_{\mathbf{f}}^{\mathbf{g}}, \nonumber \\
&\text{s.t.} & \sum_{\mathbf{g}\in{\cal F}^4} (\alpha_{\mathbf{f}}^{\mathbf{g}} + \beta_{\mathbf{f}}^{\mathbf{g}}) \le 1, \nonumber \\
& & \alpha_{\mathbf{f}}^{\mathbf{g}} = 0, \quad \forall (\mathbf{f},\mathbf{g}) \notin I_s, \nonumber \\
& & \beta_{\mathbf{f}}^{\mathbf{g}} = 0, \quad \forall (\mathbf{f},\mathbf{g}) \notin I_d, \nonumber \\
& & \alpha_{\mathbf{f}}^{\mathbf{g}}, \beta_{\mathbf{f}}^{\mathbf{g}} \in \{0,1\}, \quad \forall \mathbf{g}. \nonumber
\end{eqnarray}
Note that the variables are restricted to integer values $\{0,1\}$.

Since the queues form a Markov chain, we can use Foster-Lyapunov theorem (see Proposition $5.3$ in \cite{B:A03}) in order to prove the stability. We assume that $r_{\mathbf{g}} > 0$ for all $\mathbf{g}$. Otherwise, the queues at the relays corresponding to zero rates can be removed without affecting the rates supportable by the system and the stability of the system. Consider the Lyapunov function 
$$V(\underline{Q}[t]) = Q^2_s[t] + \sum_{n=1}^{2}\sum_{\mathbf{g}\in{\cal F}^4} (r_{\mathbf{g}} Q_{n}^{\mathbf{g}}[t])^2,$$ 
where $\underline{Q}[t]$ denotes the vector of all queue lengths. Let an optimal assignment to problem (\ref{back_pressure}) be ${\hat{\alpha}}_{\mathbf{f}}^{\mathbf{g}},{\hat{\beta}}_{\mathbf{f}}^{\mathbf{g}}$ and $E$ be the event $Q_s[t] + A[t] \ge \sum_{\mathbf{g}\in\mathcal{F}^4}r_{\mathbf{g}}{\hat{\alpha}}_{\mathbf{f}}^{\mathbf{g}}$. We have
\begin{eqnarray}
Q^2_s[t+1] &= &\left(Q_s[t] + A[t] - \left(\sum_{\mathbf{g}\in\mathcal{F}^4}r_{\mathbf{g}}{\hat{\alpha}}_{\mathbf{f}}^{\mathbf{g}}\right)\mathbf{1}_{\{E\}}\right)^2 \nonumber \\
&\le &\left(Q_s[t] + A[t] - \sum_{\mathbf{g}\in\mathcal{F}^4}r_{\mathbf{g}}{\hat{\alpha}}_{\mathbf{f}}^{\mathbf{g}}\right)^2 \nonumber \\
&\le &Q^2_s[t] + {A}^2[t] + \left(\sum_{\mathbf{g}\in\mathcal{F}^4}r_{\mathbf{g}}{\hat{\alpha}}_{\mathbf{f}}^{\mathbf{g}}\right)^2 - \nonumber \\
& & 2Q_s[t]\left(\sum_{\mathbf{g}\in\mathcal{F}^4} r_{\mathbf{g}}{\hat{\alpha}}_{\mathbf{f}}^{\mathbf{g}} - A[t]\right). \nonumber
\end{eqnarray}

Similarly, we have
\begin{eqnarray}
\left(r_{\mathbf{g}}Q_{n}^{\mathbf{g}}[t+1]\right)^2 &\le &\left(r_{\mathbf{g}}Q_{n}^{\mathbf{g}}[t] + r_{\mathbf{g}}{\hat{\alpha}}_{\mathbf{f}}^{\mathbf{g}} - r_{\mathbf{g}}{\hat{\beta}}_{\mathbf{f}}^{\mathbf{g}}\right)^2 \nonumber \\
&= &\left(r_{\mathbf{g}}Q_{n}^{\mathbf{g}}[t]\right)^2 + r_{\mathbf{g}}^2 \left({\hat{\beta}}_{\mathbf{f}}^{\mathbf{g}}-{\hat{\alpha}}_{\mathbf{f}}^{\mathbf{g}}\right)^2 - \nonumber \\
& & 2r_{\mathbf{g}}Q_{n}^{\mathbf{g}}[t]r_{\mathbf{g}}\left({\hat{\beta}}_{\mathbf{f}}^{\mathbf{g}}-{\hat{\alpha}}_{\mathbf{f}}^{\mathbf{g}}\right). \nonumber
\end{eqnarray}

Let $a_{\mathbf{f}}^{\mathbf{g}},b_{\mathbf{f}}^{\mathbf{g}}$ be any feasible assignment to the optimization problem (\ref{max_rate_eq}). Using the law of iterated expectations (conditioning on the fading state), we obtain
\begin{eqnarray}
\lefteqn{\mathbf{E}\left[V(\underline{Q}[t+1])-V(\underline{Q}[t])|\underline{Q}[t]\right]} \nonumber \\
 &\le &\sum_{\mathbf{f}\in\mathcal{F}^4} \pi_{\mathbf{f}} \left[- 2Q_s[t] \left(\sum_{\mathbf{g}\in\mathcal{F}^4}r_{\mathbf{g}}{\hat{\alpha}}_{\mathbf{f}}^{\mathbf{g}} - \lambda\right)- \right.\nonumber \\ 
 & &\left. \sum_{n=1}^{2}\sum_{\mathbf{g}\in\mathcal{F}^4} \left( 2r_{\mathbf{g}}Q_n^{\mathbf{g}}[t]r_{\mathbf{g}}\left({\hat{\beta}}_{\mathbf{f}}^{\mathbf{g}}-{\hat{\alpha}}_{\mathbf{f}}^{\mathbf{g}}\right)\right) + M\right] \nonumber \\
 &= &2\left[Q_s[t]\left(\lambda - \sum_{\mathbf{f},\mathbf{g}\in\mathcal{F}^4} (\pi_{\mathbf{f}} r_{\mathbf{g}} {\hat{\alpha}}_{\mathbf{f}}^{\mathbf{g}})\right)+ \right.\nonumber \\ 
 & &\left.\sum_{n=1}^{2}\sum_{\mathbf{g}\in\mathcal{F}^4} r_{\mathbf{g}}Q_n^{\mathbf{g}}[t] \left(\sum_{\mathbf{f}\in\mathcal{F}^4} \left(\pi_{\mathbf{f}} r_{\mathbf{g}}({\hat{\alpha}}_{\mathbf{f}}^{\mathbf{g}}-{\hat{\beta}}_{\mathbf{f}}^{\mathbf{g}})\right)\right)\right] + M \nonumber \\
 &\le &2\left[Q_s[t]\left(\lambda - \sum_{\mathbf{f},\mathbf{g}} (\pi_{\mathbf{f}} r_{\mathbf{g}} a_{\mathbf{f}}^{\mathbf{g}})\right)+ \right. \nonumber \\
 & &\left.\sum_{n=1}^{2}\sum_{\mathbf{g}\in\mathcal{F}^4} r_{\mathbf{g}}Q_n^{\mathbf{g}}[t] \left(\sum_{\mathbf{f}\in\mathcal{F}^4} \left(\pi_{\mathbf{f}} r_{\mathbf{g}}(a_{\mathbf{f}}^{\mathbf{g}}-b_{\mathbf{f}}^{\mathbf{g}})\right)\right)\right] + M, \nonumber
 \end{eqnarray}
where $M$ is a finite value, as the variance associated with the arrival process is bounded and all $r_{\mathbf{g}}$ are finite. The last inequality holds due to the following reason: Consider the linear program (LP) obtained by relaxing the integer constraints of the optimization problem (\ref{back_pressure}). This relaxation is tight as LPs have at least one optimal solution which is a boundary point. The feasible assignment set $a_{\mathbf{f}}^{\mathbf{g}},b_{\mathbf{f}}^{\mathbf{g}}$ to the optimization problem (\ref{max_rate_eq}) is a subset of the feasible set for the LP. 

We will now show that for $\lambda<r_{max}$, there is strict negative drift on the set of all possible queue states, except on a compact subset. For a given state $\mathbf{g}$, if $\sum_{(\mathbf{f},\mathbf{g})\in I_d} \pi_{\mathbf{f}} = 0$, then $Q_n^{\mathbf{g}}[t]=0$ for all $t,n$ as the algorithm will never choose to transmit to these queues. Let $\phi = \frac{\lambda}{r_{max}}$. Since $\lambda<r_{max}$, $\phi < 1$. Let $\hat{a}_{\mathbf{f}}^{\mathbf{g}}, \hat{b}_{\mathbf{f}}^{\mathbf{g}}$ be an optimal solution to the problem (\ref{max_rate_eq}). Note that the following is another trivial feasible assignment for the problem (\ref{max_rate_eq}): $a_{\mathbf{f}}^{\mathbf{g}}=0$, $b_{\mathbf{f}}^{\mathbf{g}}=\frac{1}{|\mathcal{F}^2|}$ for $(\mathbf{f},\mathbf{g})\in I_d$, $b_{\mathbf{f}}^{\mathbf{g}}=0$ for $(\mathbf{f},\mathbf{g})\notin I_d$. Since the feasible set is convex, any convex combination is also feasible. In particular consider the following convex combination: $\bar{\alpha}_{\mathbf{f}}^{\mathbf{g}}=\frac{(\phi+1)}{2}\hat{\alpha}_{\mathbf{f}}^{\mathbf{g}}$, $\bar{\beta}_{\mathbf{f}}^{\mathbf{g}} = \frac{(\phi+1)}{2}\hat{\beta}_{\mathbf{f}}^{\mathbf{g}}+\frac{(1-\phi)}{2}\frac{1}{|\mathcal{F}^2|}\mathbf{1}_{\{(\mathbf{f},\mathbf{g})\in I_d\}}$. For this assignment, it is clear that $(\lambda - \sum_{\mathbf{f},\mathbf{g}\in\mathcal{F}^4} \pi_{\mathbf{f}}r_{\mathbf{g}}\bar{\alpha}_{\mathbf{f}}^{\mathbf{g}} )< 0$, and, for any given state $\mathbf{g}$, $\sum_{\mathbf{f}\in\mathcal{F}^4} (\pi_{\mathbf{f}} r_{\mathbf{g}}(\bar{\alpha}_{\mathbf{f}}^{\mathbf{g}}-\bar{\beta}_{\mathbf{f}}^{\mathbf{g}})) <0$ if $\sum_{(\mathbf{f},\mathbf{g})\in I_d} \pi_{\mathbf{f}} > 0$. This completes the proof.

\end{document}